\documentclass[10pt,preprint2]{aastex}
\onecolumn
\usepackage{bm}
\usepackage{dcolumn}
\usepackage{graphicx}
\usepackage{latexsym,color}
\usepackage{amssymb}
\usepackage{amsmath,amssymb,amsthm,mathrsfs,amsfonts,dsfont}
\usepackage{color}
\usepackage[hyperfootnotes=false]{hyperref}
\usepackage{enumitem}

\def\be{\begin{equation}}
\def\ee{\end{equation}}
\def\bea{\begin{eqnarray}}
\def\eea{\end{eqnarray}}

\newcommand{\non}[1]{{\LARGE{\not}}{#1}}
\newcommand{\il}{~}

\makeatletter
\shorttitle{Ringed  accretion disks}
\shortauthors{D. Pugliese and Z. Stuchl\'{\i}k}

\begin{document}

\title{Ringed accretion disks: instabilities}
\author{D. Pugliese and Z. Stuchl\'{\i}k}
\affil{
Institute of Physics  and Research Centre of Theoretical Physics and Astrophysics, Faculty of Philosophy \& Science,
  Silesian University in Opava,
 Bezru\v{c}ovo n\'{a}m\v{e}st\'{i} 13, CZ-74601 Opava, Czech Republic
}
\email{d.pugliese.physics@gmail.com;zdenek.stuchlik@physics.cz}
\begin{abstract}
We analyze the possibility that several  instability points may be  formed, due to the Paczy{\'n}ski  mechanism of violation of mechanical equilibrium,  in the orbiting matter   around a supermassive Kerr black hole. We consider   recently proposed model of  ringed accretion disk,  made up   by several tori (rings) which can be  corotating  or counterrotating  relative to the   Kerr attractor {due to the history of the accretion process}. Each torus  is governed by the   general relativistic hydrodynamic Boyer  condition  of equilibrium configurations  of rotating perfect fluids.
We prove that the number of the instability points   is generally limited and depends on  the  dimensionless  spin  of the rotating   attractor.
\end{abstract}
%
%
\maketitle
  \newcommand{\downmapsto}{\rotatebox[origin=c]{-90}{$\small\mathbf{\mapsto}$}\mkern2mu}
  \newcommand{\upmapsto}{\rotatebox[origin=c]{90}{$\small\mathbf{\mapsto}$}\mkern2mu}
\newcommand{\pp}{\textbf{()}}
\newcommand{\Tem}{T^{\rm{em}}}
\section{Introduction}\label{Sec:I-introd}
The physics of accretion disks around very compact objects  as supermassive black holes (BH)  in active galactic nuclei (AGN) or quasars is  the basis of the phenomenology of the high energy astrophysics.
The   instability phases of the orbiting  toroidal configurations are particularly relevant    during the  accretion phase onto the black hole and launching of jets of matter and radiation.
On one side an important issue  is      description of the  mechanisms by which jets are launched and formed,  on the other side, the relation between  material in  accretion  and the  jet power has to be clarified.

An important subject  of investigation concerns the  dynamics of the entire black hole-disk system, which could also result  in processes producing instability   of  the entire system.
It is  often assumed that the rotational energy of the attractor may have a role in the  energy extraction and conversion of matter in jet-like structures,
 when the gravitational binding energy of accreting matter is transformed into radiation.
 An intriguing issue consists in  significance  of  the gravo-hydrostatic contribution to the jet formation and structure.
Existence of a possible jet-accretion correlation  itself is subject of many theoretical and observational investigation.  Various  analysis  suggest
that a significant part of  matter going  through
the accretion point would  possibly  power or feed
the  launching  jets. Therefore, we could actually consider an accretion-ejection engine by
referring  the   accretion
disk being connected to the relativistic jets.

The quest of a jet-accretion correlation for the formation and feeding of jets,  or also  the morphological  deformation of the matter funnels
would  provide  several constraints for  any model of
 accretion and jet formation, depending  also on  the dimensionless  spin of the black hole. In fact,
 the jet formation  appears to   be strongly related  to the intrinsic rotation  of
the central attractor
\citet{Romanova}.
The role of spin and magnetic field in  accretion disks and relativistic jets is discussed for example in
\citet{McKinneyScience} while
the relation between accretion rate and jet power in x-ray luminous elliptical galaxies has been exploited in \citet{Allen:2006mh}. Creation of jets by accretion disks orbiting magnetized black holes has  been recently proposed in \citet{Stuchlik-Kolos2015}.

The observational evidences for an
accretion-disk origin for
a radio jet in an active galaxy  is discussed in \citet{NatureMa}.
For the jet-disk connection and blazar unification see
\citep{Maraschi:2002pp,Chen:2015cga,Yu:2015lqj,Zhang:2015eka}.
For the the disk-jet connection in AGN  see \citet{Sbarrato:2014uxa} and
\citet{Coughlin:2013lva}. {For the  stellar BH casefor example \citet{MSBNNW2009} and \citet{Fender:2015kta}.}

Jets from geometrically thick disk  were studied in \citep{AbramowiczSharp(1983),Sadowski:2015ena,Okuda:2004zv,Ferreira:2003yy,Lyutikov(2009)}.
Theoretical models for the production of relativistic jets of active galactic nuclei predict that jet power is governed by the spin and mass of the central super-massive black hole, as well as by the magnetic field near the event horizon \citet{Ghisellini:2014pwa},
while   no correlation  between disk  height  and jet power appears in GR-MHD simulations presented in \citet{FragileW2012}.
Focusing on the general problem of the role of the accretion  disk-jet connection one can question  if any changes in the inner part of the  disk should
produce changes in the jet launching.
For example  \citet{Miller}  suggest that  the jet is not affected by
changes in the inner radius of the accretion disk, but it   depends on some of the  disk properties, as its flux,
temperature, and ionization.
Further open question is  the role of the inner margin of the accretion disk in the jet processes, namely the role of the marginally stable circular orbit (ISCO),
as the  definition of the inner margin of the accretion disk is extensively debated \citep{Krolik:2002ae,BMP98,2010A&A...521A..15A,Agol:1999dn,Paczynski:2000tz}. Some results lead to the conclusions  that a
standard thin accretion disk that extends to the ISCO can drive
relativistic jets \citet{Miller}.

In the present article we consider the emergence of several instability points in a ringed disk\citet{ringed}. The instability points can be associated to    accretion into the black hole, or  irregular points of  critical surfaces   representing  open structures of matter funnels, the  {proto-jets} \citep{KJA78,AJS78,Sadowski:2015jaa,Lasota:2015bii,Lyutikov(2009),Madau(1988),Sikora(1981),Stuchlik:2009jv,Stu-Sla-Hle:2000:ASTRA:,Sla-Stu:2005:CLAQG:,Stu:2005:MPLA}.
We study  different kinds of gravo-hydrostatic instability points  assuming the hydrostatic pressure and gravitational effects of a Kerr black hole background   requiring a full general relativistic treatment of geometrically thick accretion disks. 
We realize the  analysis in the framework of the ringed accretion disks (or macro-structures) recently introduced in  \citet{ringed}.
The  ringed accretion disk model is  made up by several rings rotating around a
super massive Kerr black hole attractor which  could be
created in various regimes during the evolution of matter configurations around supermassive
black holes.  We start by the consideration that different parts  of the orbiting material around a super-massive spinning black hole, characterized by different specific angular momentum, may give rise to instability in different points,  inducing eventually an overall instability of the entire  ringed  orbiting structure.

 In modelling the
evolution of supermassive black holes in AGNs both corotating and counter-rotating accretion stages are
mixed\citep{Volonteri:2002vz,Carmona-Loaiza:2015fqa,Dyda:2014pia,HL14}. Therefore, the galactic nuclei containing a supermassive black
hole could be an  environment where these macro-structure can be observed.   Toroidal rings might be   formed as remnants of several accretion regimes  occurred in various phases of the black hole life {\cite{Aligetal(2013),King:2007nx,King:2006uu,Lovelace:1996kx,Gafton:2015jja}}. These sub-structures could be eventually  reanimated in non isolated systems where the central attractor is interacting with the  environment, or in some kinds of binary systems.  Some additional matter, for example, could be  supplied into the vicinity of the
central black hole due to  tidal distortion of a star\citet{natures}.
{Possible observational evidences of these configurations were also discussed in \citep{KS10,S11etal}, more generally a possible evidence of the existence of the ringed accretion disk can be inferred from the study of the optical properties of the disk \citep{StuchlikSche:2012,Stuchlik:2010zz}.}

The individual toroidal (thick disk) configurations (the rings or sub-configurations)
are here described by the purely hydrodynamic (barotropic) model.  The  Polish Doughnut (P-D) model \citet{AbraFra} is an example of a thick, opaque and super-Eddington, radiation pressure supported accretion disks cooled by advection with low viscosity \citet{Abra83}.
Each toroid of the ringed disk is assumed being  governed by the General
Relativity hydrodynamic Boyer condition of equilibrium configurations of rotating perfect fluids.
The effects of strong gravitational fields are dominant with respect to the dissipative ones
and predominant to determine the unstable phases of the system \citep{F-D-02,Igumenshchev,AbraFra,pugtot,Pac-Wii}, where
the entropy is constant along the flow.
The von Zeipel condition is verified and accordingly the surfaces of constant angular velocity $\Omega$ and of constant specific angular momentum $\ell$ coincide \citep{M.A.Abramowicz,Chakrabarti0,Chakrabarti,Zanotti:2014haa} and  the rotation law $\ell=\ell(\Omega)$ is independent of the equation of state \citep{Lei:2008ui,Abramowicz:2008bk,AJS78}.
Properties of each  tori can be then determined by an effective potential
reflecting the background Kerr geometry.
The equipotential surfaces associated with critical points identify the toroidal surfaces of the disk.
The cusped surfaces are the critical topologies associated to the  unstable  phases of the configurations.
The outflow of matter through the cusp occuers  by
the Paczy{\'n}ski-Wiita (P-W) mechanism of violation of mechanical equilibrium of the tori, i.e.  an
instability in the balance of the gravitational and inertial forces and the pressure gradients in the fluid\citet{AbraFra}.

As in \citet{ringed}  we consider the ringed disk where
the centers of all the individual tori are  coplanar,
coinciding with the equatorial plane of the Kerr attractor. This assumption of coincidence between the orbital
planes of the rings and the equatorial plane of the axisymmetric attractor, is the simplest
scenario  for  the majority of the current analytical and numerical
models  of the extended accreting matter\footnote{Off-equatorial configurations are considered in \citep{Kovar:2008ta,Cremaschini:2013jia} and  in \citep{Dogan:2015ida,NixonKing(2012),NixonKing(2012a),NixonKing(2012b),Nixon:2013qfa,Dunhill:2014oka,Nixonetal.(2011)}
 where also strongly misaligned  disks with respect to the central supermassive BH spin are considered.}.
We face particularly  the problem of location of the inner and outer edges of the toroidal configurations  and the critical points with respect to the marginally bounded and marginally stable  circular orbits.
This analysis provides the basis for the  attractor classifications.
The  rings are related    by boundary conditions dictated by the condition of not penetration of  matter and by the geometric constraint for the equilibrium configurations determined by the geometric properties of the Kerr background reflected by  its geodesic structure.
The condition of non-penetration of matter was considered in \citet{ringed} in  a very restrictive way, avoiding collisions in the macro-structure. As a consequence of this  assumption only a point of instability would be possible in the   macro-configuration.
In the present article  we weaken this request,  investigating the possibility that there may be several  P-W points for a given macro-configuration.
We distinguish   five  types of unstable couples of orbiting configurations  (\emph{states} of the macro-configurations).
 Not all these states can actually exist. We   prove that their formation and stability depends on the  dimensionless spin of the attractor, the relative rotation of the tori with respect to the attractor, and  more importantly, the relative rotation of the fluids in the ringed disk.
Considering  the possible combination of these states, we investigate also  ringed disks consisting of more than two rings  and a  possible state correlation   occurring  when  two surfaces are in contact
leading to collision phenomena and  eventually to a topological transition of  the  state and, consequently, of  the entire macro-configuration.
We  face the problem of the {state  evolution} considering the condition for  an initial couple  of configurations   (starting state) could evolve towards    a   transition of the surface topologies with emergence of instability.

This article is structured as follows:
in the first brief  descriptive part,  Sec.\il(\ref{Sec:models}) the ringed disk model is introduced and some general considerations on the instability points are discussed.
Sec.\il(\ref{Sec:IIsec}) addresses briefly the main outcomes of the   analysis focusing on the states properties, details will be provided in future studies.
  Concluding remarks can be found in Sec.\il(\ref{Sec:Open-Concl}).
\section{Rings in the Kerr spacetime}\label{Sec:models}
\begin{figure}[h!]
\begin{center}
\begin{tabular}{ll}
 \includegraphics[scale=0.3]{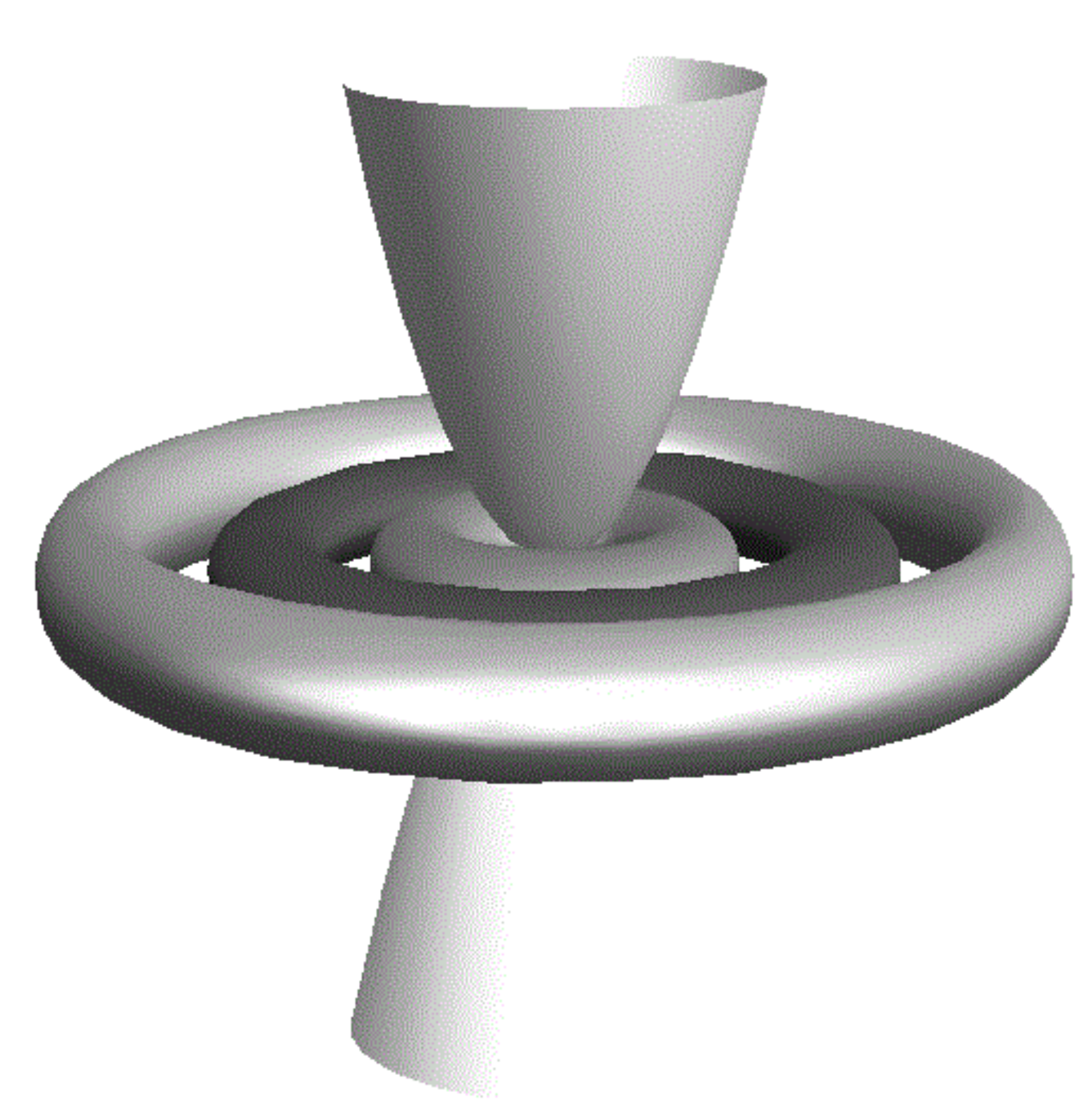}
 \includegraphics[width=.55\textwidth]{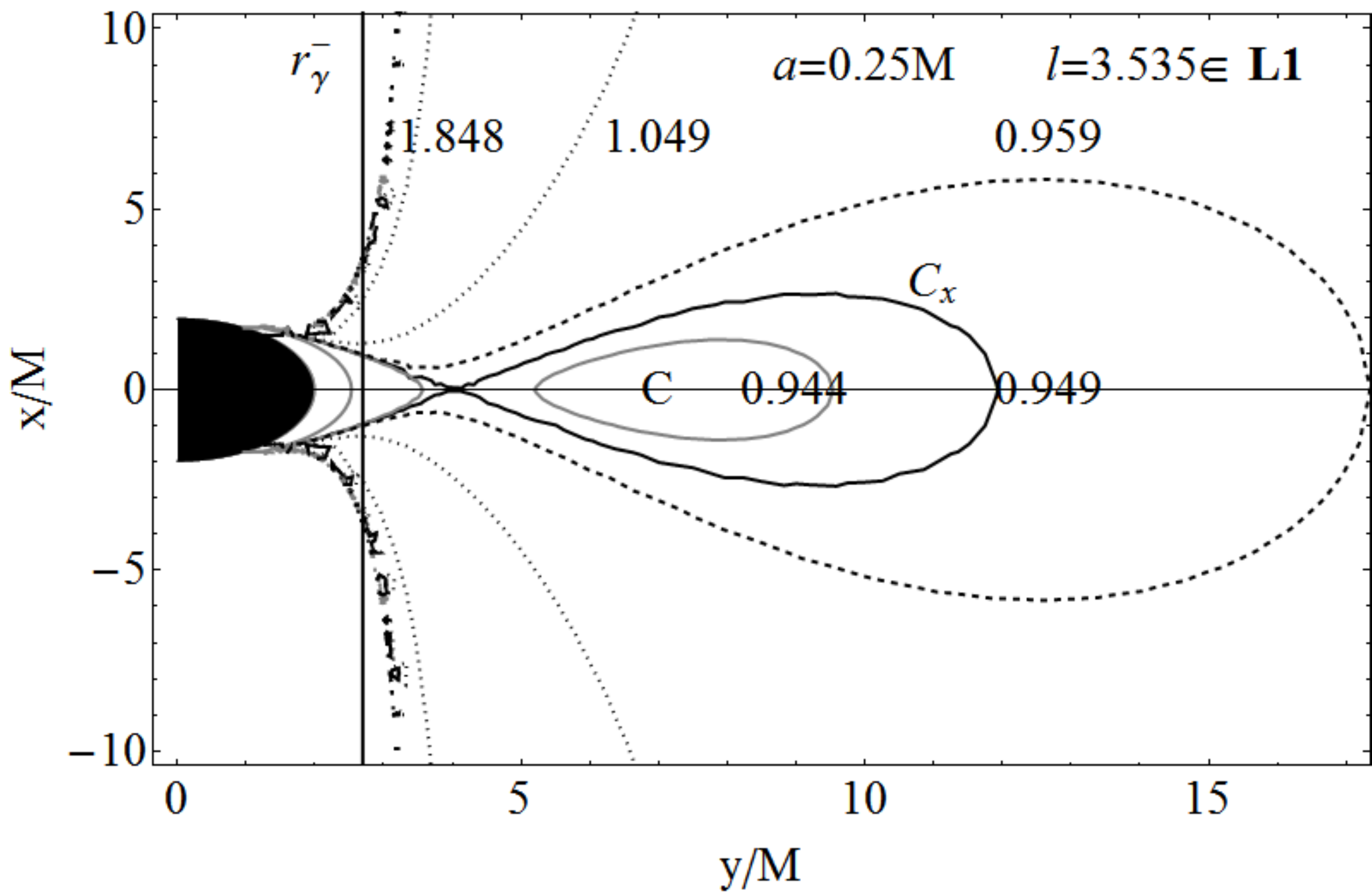}
\end{tabular}
\caption{Left: Pictorial representation of a ringed accretion disk. Right: Spacetime spin $a=0.25M$, $\ell$corotating sequences, $\ell_i\ell_j>0$, of corotating disks  $\ell_i a>0$ $\forall i j$. The outer horizon is  at $r_+=1.96825M$, the region $r<r_+$ is  black-colored. }\label{Figs:ApproxPlo}
\end{center}
\end{figure}

\begin{figure}[h!]
\begin{center}
\begin{tabular}{lr}
\includegraphics[width=0.57\textwidth]{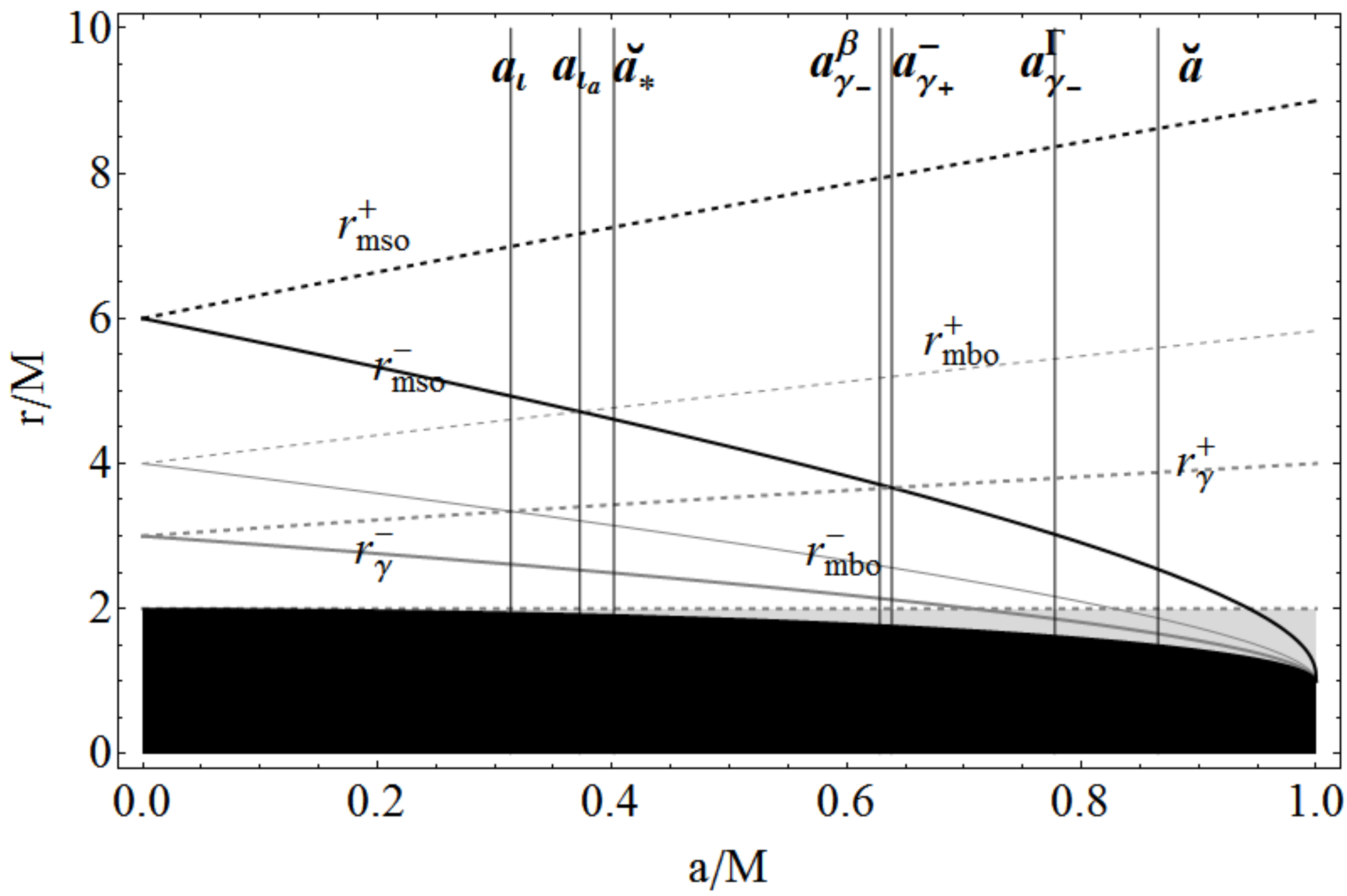}
\includegraphics[width=0.57\textwidth]{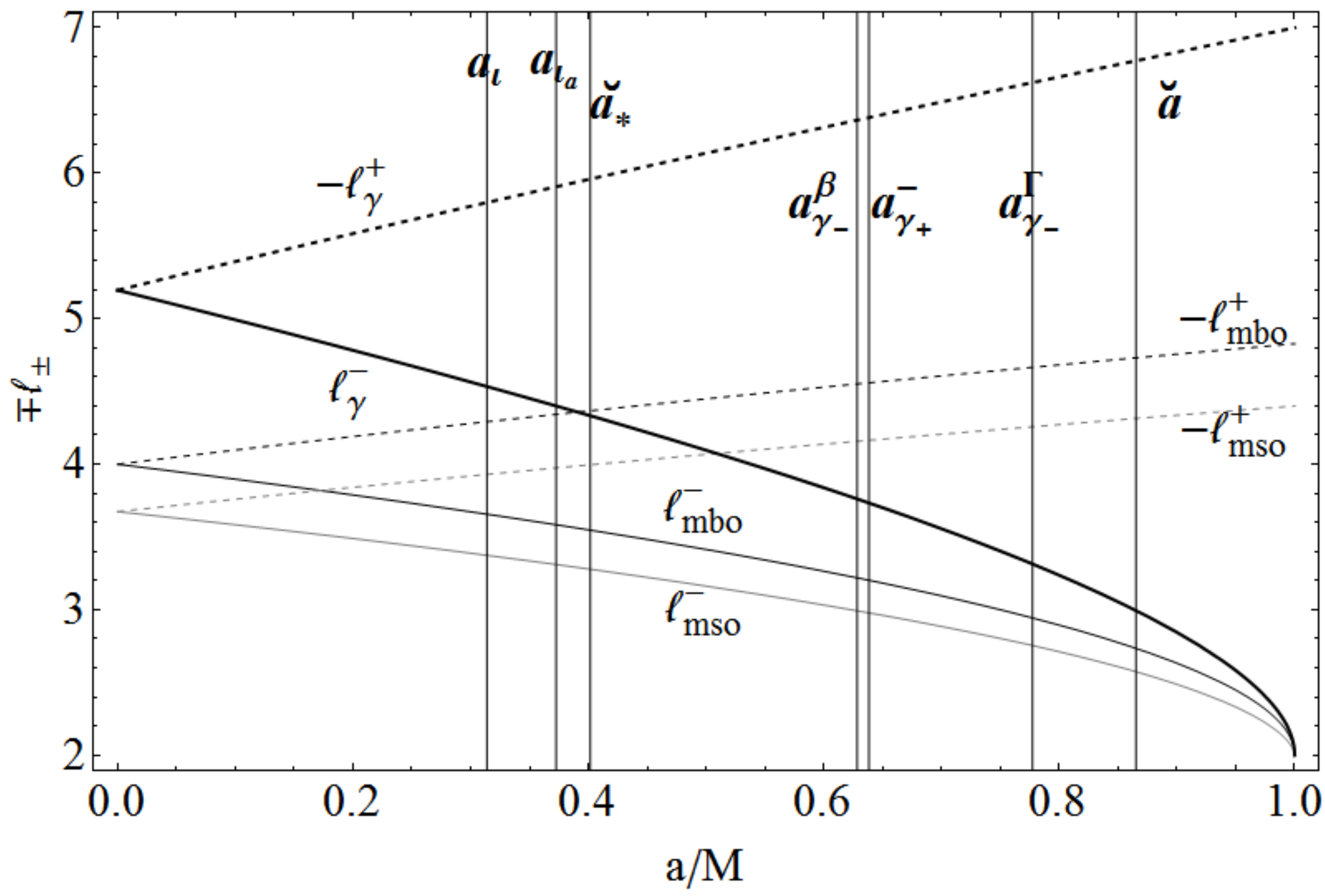}
\end{tabular}
\caption{Geodesic structure of the Kerr geometry: notable radii  $r_{\mathcal{N}}\equiv \{r_{\gamma}^{\pm}, r_{mbo}^{\pm},r_{mso}^{\pm}\}$ (left panel), and the   fluid specific angular momentum $\ell^{\pm}_i=\ell^{\pm}(r^{\pm}_i)$ (right panel)  where $r^{\pm}_i\in r_{\mathcal{N}}^{\pm}$. Some notable spacetime spin-mass ratios are also plotted. Black region is $r<r_{+}$, where $r_+$ is the outer horizon for a Kerr geometry, gray region is $]r_+, r_{\epsilon}^+]$, where $r_{\epsilon}^+=2M$ is the outer ergosurface.}\label{Figs:PlotdisolMsM}
\end{center}
\end{figure}

In this section, we  introduce the main notation  used in this work and we will make reference also to  some basic notions introduced in \citet{ringed}.
The Kerr  metric tensor  in the Boyer-Lindquist (BL)  coordinates
\( \{t,r,\theta ,\phi \}\) reads
\bea \label{alai}&& ds^2=-dt^2+\frac{\rho^2}{\Delta}dr^2+\rho^2
d\theta^2+(r^2+a^2)\sin^2\theta
d\phi^2+\frac{2M}{\rho^2}r(dt-a\sin^2\theta d\phi)^2\ ,
\\
\nonumber
&&
 \mbox{where}\quad\rho^2\equiv r^2+a^2\cos\theta^2\quad \mbox{and } \quad \Delta\equiv r^2-2 M r+a^2,
\eea
where
 the specific angular momentum is  $a=J/M\in]0,1]$, $M$ is a mass parameter and  $J$ is the
total angular momentum of the gravitational source. The extreme Kerr black hole  has dimensionless spin $a/M=1$, while  the non-rotating  limiting case $a=0$ is the   Schwarzschild metric\footnote{We adopt the
geometrical  units $c=1=G$ and  the $(-,+,+,+)$ signature, Greek indices run in $\{0,1,2,3\}$.  The   four-velocity  satisfy $u^{\alpha} u_{\alpha}=-1$. The radius $r$ has unit of
mass $[M]$, and the angular momentum  units of $[M]^2$, the velocities  $[u^t]=[u^r]=1$
and $[u^{\varphi}]=[u^{\theta}]=[M]^{-1}$ with $[u^{\varphi}/u^{t}]=[M]^{-1}$ and
$[u_{\varphi}/u_{t}]=[M]$. For the seek of convenience, we always consider the
dimensionless  energy and effective potential $[V_{eff}]=1$ and an angular momentum per
unit of mass $[L]/[M]=[M]$.}.
The horizons $r_-<r_+$ and the outer static limit $r_{\epsilon}^+$ are respectively given by:
\bea
r_{\pm}\equiv M\pm\sqrt{M^2-a^2};\quad r_{\epsilon}^{+}\equiv M+\sqrt{M^2- a^2 \cos\theta^2};
\eea
where $r_+<r_{\epsilon}^+$ on   $\theta\neq0$  and  $r_{\epsilon}^+=2M$  in the equatorial plane $\theta=\pi/2$.
We  consider   toroidal  configurations of  perfect fluids orbiting a    Kerr black hole \textbf{(BH)} attractor.
Importantly, the Kerr  geometry has two Killing vectors:    $\xi_{\phi}=\partial_{\phi} $, rotational Killing field,  and  $\xi_{t}=\partial_{t} $,  which is
the Killing field representing the stationarity of the  spacetime. Line element (\ref{alai}) is  indeed independent of $\phi$ and $t$. Consequently,
the quantities
\be\label{Eq:after}
{E} \equiv -g_{\alpha \beta}\xi_{t}^{\alpha} p^{\beta}=-p_t,\quad L \equiv
g_{\alpha \beta}\xi_{\phi}^{\alpha}p^{\beta}=p_{\phi}\ ,
\ee
are  constants of motion
 and $p_a$  is the particle four--momentum.
 We  can  limit the  analysis of the test particle circular motion to the case of  positive values of $a$
for corotating  $(L>0)$ and counterrotating   $(L<0)$ orbits with respect to the black hole. In fact the
 metric tensor (\ref{alai}) is  invariant under the application of any two different transformations: $x^\alpha\rightarrow-x^\alpha$
  for one of the coordinates $(t,\phi)$, or the metric parameter $a$ and, consequently,    the    test particle dynamics is invariant under the mutual transformation of the parameters
$(a,L)\rightarrow(-a,-L)$.
The constant $L$ in Eq.\il(\ref{Eq:after}) may be interpreted       as the axial component of the angular momentum  of a   particle for
timelike geodesics and $E$ as representing the total energy of the test particle
 coming from radial infinity, as measured  by  a static observer at infinity.
In this work we deal with a  one-species particle perfect  fluid system, which is   described by  the  energy momentum tensor
\be\label{E:Tm}
T_{\alpha \beta}=(\varrho +p) u_{\alpha} u_{\beta}+\  p g_{\alpha \beta},
\ee
where $\varrho$ and $p$ are  the total energy density and
pressure, respectively, as measured by an observer comoving with the fluid whose four-velocity $u^{\alpha}$  is
a timelike flow vector field.
The  fluid dynamics  is described by the \emph{continuity  equation} and the \emph{Euler equation} respectively:
\bea\label{E:1a0}
u^\alpha\nabla_\alpha\varrho+(p+\varrho)\nabla^\alpha u_\alpha=0\, ,\quad
(p+\varrho)u^\alpha \nabla_\alpha u^\gamma+ \ h^{\beta\gamma}\nabla_\beta p=0\, ,
\eea
where  $\nabla_\alpha g_{\beta\gamma}=0$ and
for the
symmetries of the problem, we  assume $\partial_t \mathbf{Q}=0$ and
$\partial_{\varphi} \mathbf{Q}=0$,  with $\mathbf{Q}$ being a generic spacetime tensor.
In Eq.\il(\ref{E:1a0}), $h_{\alpha \beta}=g_{\alpha \beta}+ u_\alpha u_\beta$ is  the projection tensor  \citep{Pugliese:2011aa,pugtot}.
Assuming a barotropic equation of state $p=p(\varrho)$, we investigate  the  fluid toroidal configurations (with  $u^{\theta}=0$) centered on  the  plane $\theta=\pi/2$, and  defined by the constraint
$u^r=0$.
Thus, in this setup
we find from the Euler  equation (\ref{E:1a0})
\be\label{Eq:scond-d}
\frac{\partial_{\mu}p}{\varrho+p}=-{\partial_{\mu }W}+\frac{\Omega \partial_{\mu}\ell}{1-\Omega \ell},\quad \ell\equiv \frac{L}{E},\quad W\equiv\ln V_{eff}(\ell),\quad V_{eff}(\ell)=u_t= \pm\sqrt{\frac{g_{\phi t}^2-g_{tt} g_{\phi \phi}}{g_{\phi \phi}+2 \ell g_{\phi t} +\ell^2g_{tt}}},
\ee
(the  continuity equation  in Eq.\il(\ref{E:1a0})
is  identically satisfied as consequence of the applied conditions and symmetries). The function $V_{eff}(\ell)$ is  the effective potential for the fluid, the function $W$ is the Paczy{\'n}ski-Wiita  (P-W) potential, reflecting the background  Kerr geometry and the centrifugal effects, and we assume here a  constant  and conserved   specific angular momentum $\ell$  (see also \citep{Lei:2008ui,Abramowicz:2008bk}). Finally,
  $\Omega$ is the relativistic angular frequency of the fluid relative to the distant observer.
Similarly to the case of the test particle dynamics,
the  function  $V_{eff}(\ell)$  in Eq.\il(\ref{Eq:scond-d})  is invariant under the mutual transformation of  the parameters
$(a,\ell)\rightarrow(-a,-\ell)$. Therefore, we can limit the analysis to  positive values of $a>0$,
for \emph{corotating}  $(\ell>0)$ and \emph{counterrotating}   $(\ell<0)$ fluids and    we adopt the notation $(\pm)$  for  counterrotating or corotating matter  respectively.
We  consider a fully general relativistic model of ringed accretion disk formed by several  corotating and counterrotating toroidal rings orbiting a supermassive Kerr attractor \citep{ringed}.
In a given spacetime characterized by  the dimensionless spin $(a/M)$,  each toroid of the ringed  disk is governed by the   General Relativity
   hydrodynamic Boyer  condition  of equilibrium configurations  of rotating perfect fluids.
According to the  Boyer theory on the equipressure surfaces applied to a  P-D  torus,
the toroidal surfaces  are the equipotential surfaces of the effective potential  $V_{eff}(\ell,r)$, solutions of  $V_{eff}=K=$constant  or $ \ln(V_{eff})=\rm{c}=\rm{constant}$   \citep{Boy:1965:PCPS:,Raine}.  These  correspond also to the surfaces of constant density, specific angular momentum $\ell$, and constant  relativistic angular frequency  $\Omega$, where $\Omega=\Omega(\ell)$  as a consequence of the von Zeipel theorem \citep{M.A.Abramowicz,Zanotti:2014haa}.
Each Boyer surface is  uniquely identified by the couple of parameters $\mathbf{p}\equiv (\ell,K)$.
Since the  toroidal configuration  can be corotating,  $\ell a>0 $, or counterrotating,   $\ell a<0$, with respect to the black hole  rotation $(a>0)$,  assuming    first a couple $(C_a, C_b)$ with specific angular momentum $(\ell_a, \ell_b)$,  orbiting  in   the equatorial plane of a given Kerr \textbf{BH},   we need to introduce   the concept  of
 \emph{$\ell$corotating} disks,  defined by  the condition $\ell_{a}\ell_{b}>0$, and \emph{$\ell$counterrotating} disks defined  by the relations   $\ell_{a}\ell_{b}<0$.  The two $\ell$corotating tori  can be both corotating, $\ell a>0$, or counterrotating,  $\ell a<0$, with respect to the central attractor.
Fig.\il(\ref{Figs:ApproxPlo})
represents a ringed accretion disk with equilibrium configurations and   open ($O_x$ proto-jet) surfaces.
It will be important to consider in the analysis of the ringed disks  the   notable  radii  $r_{\mathcal{N}}^{\pm}\equiv \{r_{\gamma}^{\pm}, r_{mbo}^{\pm},r_{mso}^{\pm}\}$,  defining   the geodesic structure of the Kerr spacetime  with respect to  the matter distribution:  a geometric property of the spacetime    consisting of the   union of the   orbital regions with boundaries in $r_{\mathcal{N}}$--as sketched in Fig.\il(\ref{Figs:PlotdisolMsM}).
It can be decomposed, for $a\neq0$, into   $r_{\mathcal{N}}^-$ for the corotating and   $r_{\mathcal{N}}^+$  counterrotating matter.  
Since the intersection of $r_{\mathcal{N}}^{\pm}$ is not empty, the analysis of the geodesic structure  will be particularly  relevant in  the characterization of the   $\ell$counterrotating sequences\citep{ringed}.
Specifically, for timelike particle orbits, $r_{\gamma}^{\pm}$ is  the \emph{marginally circular orbit}  or  the photon circular orbit, timelike  circular orbits  can fill  the spacetime region $r>r_{\gamma}^{\pm}$. The \emph{marginally stable circular orbit}  $r_{mso}^{\pm}$: stable orbits are in $r>r_{mso}^{\pm}$ for counterrotating and corotating particles respectively.  The \emph{marginally  bounded circular  orbit}  is $r_{mbo}^{\pm}$, where
 $E_{\pm}(r_{mbo}^{\pm})=1$  \citep{Pugliese:2011xn,Pugliese:2013zma,ergon}.
 Given $r_i\in \mathcal{R}$,  we adopt  the  notation for any function $\mathbf{Q}(r):\;\mathbf{Q}_i\equiv\mathbf{Q}(r_i)$, therefore for example $\ell_{mso}^+\equiv\ell_+(r_{mso}^+)$, and more generally given the radius  $r_{\ast}$ and the function  $\mathbf{Q}(r)$,  there is $\mathbf{Q}_{\ast}\equiv\mathbf{Q}(r_{\ast})$.
We focus on the solution of Eq.\il(\ref{Eq:scond-d}), $W=$constant, associated to the critical points of the effective potential, with constant angular momentum and parameter $K$.
Thus, we  consider the orbital region $\Delta r_{crit}\equiv[r_{Max}, r_{min}]$, whose boundaries correspond to the  maximum and minimum points of the effective potential respectively.
The inner edge  of the Boyer surface  must be   at $r_{in}\in\Delta r_{crit}$,
the outer edge  is at $r_{out}>r_{min}$.   A further  matter configuration   closest to the black hole is  at $r_{in}<r_{max}$. 
 The limiting case of $K_{\pm}=K_{min}^{\pm}$ corresponds to a one-dimensional ring of matter  located in  $r_{min}^{\pm}$.
The centers $r_{cent}$  of the closed configurations $C_{\pm}$ are located at the minimum points  $r_{min}>r_{mso}^{\pm}$  of the effective potential, where the hydrostatic pressure reaches a  maximum. The toroidal surfaces are characterized by $K_{\pm}\in [K^{\pm}_{min}, K^{\pm}_{Max}[ \subset]K_{mso}^{\pm},1[\equiv \mathbf{K0}$ and  momentum $\ell_{\pm}\lessgtr\ell_{mso}^{\pm}\lessgtr0$  respectively.
The maximum points of the effective potential $r_{Max}$  correspond to minimum points of the hydrostatic pressure and the P-W  points  of  gravitational and hydrostatic instability.
An  accretion  overflow of matter from the  closed, cusped  configurations in   $C^{\pm}_x$ (see Fig.\il(\ref{Figs:ApproxPlo}) towards the attractor  can occur from the instability point  $r^{\pm}_x\equiv r_{Max}\in]r_{mbo}^{\pm},r_{mso}^{\pm}[$, if $K_{Max}\in \mathbf{K0}^{\pm}$  with specific angular momentum $\ell\in]\ell_{mbo}^+,\ell_{mso}^+[\equiv\mathbf{L1}^+$  or $\ell\in]\ell_{mso}^-,\ell_{mbo}^-[\equiv \mathbf{L1}^-$. Otherwise,  there can be  funnels of  material along an open configuration   $O^{\pm}_x$,  proto-jets or for brevity jets, which are limiting topologies for the  closed  surfaces \citep{KJA78,AJS78,Sadowski:2015jaa,Lasota:2015bii,Lyutikov(2009),Madau(1988),Sikora(1981)}
 with   $K^{\pm}_{Max}\geq1$ ($\mathbf{K1}^{\pm}$), ``launched'' from the point $r^{\pm}_{J}\equiv r_{Max}\in]r_{\gamma}^{\pm},r_{{mbo}}^{\pm}]$ with specific angular momentum $\ell\in ]\ell_{\gamma}^+,\ell_{mbo}^+[\equiv \mathbf{L2}^+ $ or $]\ell_{mbo}^-,
  \ell_{\gamma}^-[\equiv\mathbf{L2}^-$.
  Equilibrium configurations, with topology $C$,  exist for $\pm\ell_{\mp}>\pm\ell_{mso}^{\mp}$ centered in  $r>r_{mso}^{\mp}$ respectively;  no  maxima of the effective potential exist for $\pm\ell_{\mp}>\ell_{\gamma}^{\pm}$ ($\mathbf{L3}^{\mp}$), and therefore only equilibrium configurations are possible.
In general, we   mean by the  label $(i)$ with $i\in\{1,2,3\}$ respectively, any  quantity $\mathbf{Q}$ relative to the range  of specific angular momentum $\mathbf{Li}$ respectively;  for example,
$C_2^+$ indicates a closed  regular counterrotating configuration with specific angular momentum  $\ell_2^+\in\mathbf{L2}^+$.
A relevant aspect  in this model  is that the presence of critical  points  is regulated mainly by geometric factors \citep{pugtot,Pugliese:2012ub}.
Following the discussion in \citet{ringed} for the unstable modes of a   ringed tori, we  consider here two types of instabilities emerging in an orbiting macro-structure,   leading  to a global instability  in the ringed disk.
Correspondingly,  we can  distinguish two distinct models  of unstable ringed torus $\mathbf{C}^n=\bigcup^nC_i$ made up by $n$ sub-configurations  (\emph{order} $n$)  with degenerate topology :  the macro-structure
$\mathbf{C_{\odot}^n}$, with the number $\mathfrak{r}\in[0,n-1]$ (\emph{rank}  of the
$\mathbf{C_{\odot}^n}$ torus) of contact points between the boundaries of two consecutive rings,   and the macro-structure $\mathbf{C^n_x}$, with    $\mathfrak{r_x}\in[0,n]$  instability P-W points. The number  $\mathfrak{r_x}$ is called \emph{rank} of the ringed disk  $\mathbf{C^n_x}$. Finally, we have  the macro-structure $\mathbf{C_{\odot}^x}^n$, characterized   at last by one  contact  point that is also an instability point.
The P-W  local  instability affects  one or more    rings of the ringed disk  decomposition, and  then it can  destabilize the whole ringed disk in the initial    $\mathbf{C}_x^n$ topology for collisions of matter between the unstable ring and its consecutive sub-configurations,   resulting eventually  in a different  topology  of  the entire macro-structure, when the rings are   no more separated and a feeding (overlapping)   of material  occurs.
As proved  in \citet{ringed}, if    $\mathfrak{r}_x=1$  and     the inner ring $\mathbf{C}_x^1$   of its decomposition is in accretion, then  the whole  ringed disk could  be globally stable if  the outer edge of the accreting ring satisfies  proper conditions on  the fluid specific angular momentum.

In this article, we consider  the existence  of several instability points     investigating  the $\mathbf{C}_x^n$  models with rank $r_x\in[1,n]$,
  which could also include  the  open and cusped  $O_x$ sub-configurations and the situations where  two rings may be in contact
or  \emph{geometrically correlated }  according to constraints settled on their morphology or topology evolution and the geodesic structure of the spacetime.
 A contact in this model causes  collision and  penetration of matter, eventually with the feeding of one sub-configuration with material and supply of specific angular momentum of  another consecutive  ring of the decomposition. This mechanism  could possibly end  in a change of   the ring disk  morphology and topology.
In the  following sections we will face the analysis of the $n$ order decomposition   starting from the characterization of the  configurations of the order $n=2$. These can be  regarded as {seeds} for the construction of
higher-order macro-configurations  generated from  this initial couple.
We define as \emph{state}  or main state a ringed disk of the  order $n=2$ with fixed topology.
The notion of state  is    useful to clarify different aspects of the  macro-configuration structure and evolution and to deal with the many different  cases occurring even for one couple of rings. {In fact, a  ringed disk of the  order  $n=2$, with fixed  equal critical topology   could  be in $n=8$ different states according to their rotation and   relative position of the centers: $n=4$ different states, if  the rings are $\ell$corotanting,  and $n=4$ for  $\ell$countorrotating rings, considering also the relative location of points of minimum pressure.  Then, the couple $({C_x,O_x})$, with different but fixed topology,  could  be in $n=16$ different states.
 For a main state  $C_i-\pp_x$,   considering  also  the class of magnitude  $\mathbf{Li}$  where $i=\{1,2,3\}$ for the equilibrium configuration,  we need to address  $n=48$ different states. As a result of our analysis we are able to reduce  all these possible   states to the ones listed in Table\il(\ref{Table:commetime}), whereas details on the restrictions for combinations of different states in ringed disks made up by more then  two rings will be presented in future study\cite{coop}.}
   We will consider therefore
{five} main states, with at last one critical topology determining   the constraints on the state existence and {evolution}.
We will prove   that not all  the initial and final states are possible and not all the evolutions are actually possible in  all spacetimes.
\section{States and   macro-configuration}\label{Sec:IIsec}
To simplify the presentation of  the results, we  summarize in this section comments on Tables (\ref{Table:commetime}), (\ref{Table:commest}) and  (\ref{Table:commestmp}) which  schematically introduce some main outcomes of this  analysis,
whereas
 we specify the different states   for the
 $\ell$corotating couples  in \citet{coop} where  we extend  our investigation to the multiple configurations of the decomposition of  the order $n>2$
 providing also evidence of the results and  the detailed comments on each case.
First we point out that
\be\label{Eq:spe-ci}
\mbox{ if  }\quad \pp_i<\pp_o, \quad \mbox{ where}\quad \ell_i\ell_o>0,\quad\mbox{ then}\quad \pp_i\succ\pp_o, \quad \mbox{where}\quad \pp\in\{C, C_x, O_x\}.
\ee
 For the \emph{ordered} sequences  of surface, with the notation $<$ or $>$,  we intend  the ordered sequence of  maximum points of the pressure, or $r_{min}$, minimum of the effective potential and the configuration centers. In relation to a couple  of rings,   the terms ``internal'' (inner) or ``external'' (outer), will always refer, unless otherwise specified,  to the   sequence ordered according to the center location.
If   ${C}_{i}<{C}_j$ then, for  $i<j$,  ${C}_i$ is the inner ring, the closest to attractor with  respect to ${C}_j$, and  there is    $r_{cent}^i\equiv r_{min}^{i}<r_{min}^j\equiv r_{cent}^j$. Within these definitions, the rings  $(C_i, C_{i+1})$ and $(C_{i-1}, C_{i})$ are \emph{consecutive} as  $C_{i-1}<C_i<C_{i+1}$\citep{ringed}.
The
symbols $\succ$ and $\prec$  refer instead to the sequentiality according to the  location of the \emph{minimum} points of the pressure,   or $r_{Max}$, maximum point  of the effective potential (in $\mathbf{L1}$ or $\mathbf{L2}$). 
Thus, for the  $\ell$corotanting sequence there is $\pp_i<\pp_o$ and $\pp_i\succ\pp_o$.
Then, the   largest is the magnitude of the fluid specific angular momentum and the  largest  is the  radius of the  maximum  pressure point and   more stretched on the  equatorial plane is  the configuration, the nearest to the \textbf{BH} is the instability point.
The situation for  a   $\ell$counterrotating  couple with a   critical configuration  is  determined by the two families of the notable radii $r_{\mathcal{N}}^{\pm}$   and by the associated specific angular  momenta $\ell_{\mathcal{N}}^{\pm}$--see Figs.t\il(\ref{Figs:PlotdisolMsM}). The discussion of this case  is more articulated  than the $\ell$corotating case.
 Table\il(\ref{Table:commetime}) summarizes these results.
\begin{table*}[h!]
\centering
\resizebox{1\textwidth}{!}{%
\begin{tabular}{ll}
\textbf{ $\ell$corotating  configurations}&\textbf{ $\ell$counterrotating  configurations
}\\\hline
$O_x^i<O_x^o$ and $ O_x^i\succ! O_x^o$ for  $\pm\ell_i^{\mp}<\pm\ell_o^{\mp}$&$ O_x^+\succ! O_x^-$  for  $a\in {\mathbf{A}}_{\iota}^>$- ${\mathrm{\mathbf{(\non{C})}}}$;\quad
$O_x^+\succ O_x^-$ or $O_x^+\prec O_x^-$ for  $a\in {\mathbf{A}}_{\iota}^<$
\vspace{.2cm}\\
$C_x^i<!O_x^o$ and $ C_x^i\succ! O_x^o$ for $\pm\ell_i^{\mp}<\pm\ell_o^{\mp}$ &$C_x^+\succ! O_x^- $
\vspace{.2cm}\\
$C_i<!O_x^o$ and $ O_x^o\prec! C_i$ for $\pm\ell_i^{\mp}<\pm\ell_o^{\mp}$&$C_x^-\prec! O_x^+$ for $a\in \mathbf{A}_{\iota_a}^>$-$\mathrm{\mathbf{(\non{C}_*)}}$;\;$C_x^-\non{\prec} O_x^+$ for $a\in\mathbf{A}_{\iota_a}^<$
\vspace{.2cm}\\
$C_x^a<!C_b$ for $
\pm\ell_a^{\mp}<\pm\ell_b^{\mp}$ and $\ell_a \in \mathbf{L1} \; \ell_b\in \mathbf{Li}$
&$C_x^-\prec! C_x^+$ and $ C_x^-<! C_x^+ $
\vspace{.2cm}\\
 &$C_x^-<!C_+\quad r_{x}^{-}\leq r_{in}^{+}$ for $a\in\mathbf{A}_{\iota_a}^<$-$a\in\mathbf{A}_{\iota_a}^>$-$
 \mathrm{\mathbf{(\non{C}_*)}}$
\vspace{.2cm}\\
&$C_-<C_x^+ \quad r_{x}^+\geq r_{out}^{-}$ and
$C_x^+<C_-\quad r_{in}^{-}\geq r_{out}^+$
\vspace{.2cm}\\
&$C_+<O_x^-\quad r_{out}^+\non{\leq} r_{J}^- \quad r_{in}^+\geq r_{J}^-$-$\mathrm{\mathbf{(\non{C})}}$;
$O_x^-<C_+\quad r_{out}^+\non{\leq }r_{J}^- \quad r_{in}^+\geq r_{J}^-$-$\mathrm{\mathbf{(\non{C})}}$
\vspace{.2cm}\\
&$C_-<O_x^+\quad r_{out}^-\leq r_{J}^+ \quad r_{in}^-\geq r_{J}^+$;
$ O_x^+<C_-\quad r_{out}^-\non{\leq} r_{J}^+ \quad  r_{in}^-\geq r_{J}^+$-$\mathrm{\mathbf{(\non{C}_*)}}$
\\
\hline
\end{tabular}}
\caption{Couples  with at least one  cusped  topology.  The location of the disk edges is also fixed. The  $(\non{\mathrm{\mathbf{C}}})$ is for non-correlated configurations, or $(\non{\mathrm{\mathbf{C}}}_*)$    with particularly restrictive conditions to be satisfied for a correlation to occur--see\citet{coop}.
}\label{Table:commetime}
\end{table*}
In order to prove these results
we   addressed  the  issue of  the  location of  the matter distribution $\pp_{\pm}$  with respect to  notable radii the $r_{\mathcal{N}}$.
This is in fact important particularly in the determination of a possible correlation  between  rings. We discuss  the   case of $\ell$corotating matter,
 investigating the inclusion\footnote{The \emph{inclusion} notation, $(\in, \non{\in})$ and $\in!$, will be  widely used in  Tables\il(\ref{Table:commest},\ref{Table:commestmp}). The use of $\bar{r}\in\pp$,  for the radius $\bar{r}$  and any  surface $\pp$, means that  there  can be  found  proper $K$ or $\ell$ parameters  such  that this  property is satisfied. The symbol $\in\;!$ is  a reinforcement of this inclusion, indicating that this is  a necessary relation which is  \emph{always} satisfied.  The symbol  $\non{\in}$ (meaning non-inclusion)  does not generally have any  intensifier $(!)$, which is also used in Table\il(\ref{Table:commetime}), as this analysis is to underline the  possibility of inclusion and the condition for this to be satisfied.}
 of  $r_{\mathcal{N}}^{\pm} \in \pp_{\pm}$  and of  $r_{\mathcal{N}}^{\pm} \in \pp_{\mp}$.
It is worth noting here that this  investigation   actually matches  the  broader  problematic of    the location of
the inner edge of the disk--see \citep{Krolik:2002ae,BMP98,2010A&A...521A..15A,Agol:1999dn,Paczynski:2000tz}.
Indeed, this investigation will often imply, especially for  the inclusion
$r_{\mathcal{N}}^{\pm} \in \pp_{\pm}$, a  discussion of the  location   of  these radii with respect to the inner  margin of the disk, while the location of the outer edge turns to be  important especially for the discussion of the  $r_{\mathcal{N}}^{\pm} \in \pp_{\mp}$ case. Note that the accretion or proto-jet instability  of any sub-configuration is  essentially sorted at the inner edge of each  ring. Conversely, in the accretion ringed disks, an accretion point or launching point of proto-jet can emerge  at any inner margin of any of its rings. Therefore, contrary to the common scenario for a single thick accretion disk, such points of instabilities can  emerge  in the middle of the macro-structure. 
Tables\il(\ref{Table:commest}) and (\ref{Table:commestmp}) summarize the main results of this analysis.
\begin{table*}[h!]
\centering
\begin{tabular}{lll}
\textbf{Location of the marginally bounded orbits $r_{mbo}$
}\\\hline
$r_{mbo}^{\pm}\non{\in} C^{\pm}_3$; \hspace{.5cm} $r_{mbo}^{\pm}\non{\in} C^{\pm}_2$ $r_{mbo}^{\pm}{\in}! O_x^{2_{\pm}}$; \hspace{.5cm}  $r_{mbo}^{\pm}\non{\in} C^{\pm}_1$ $r_{mbo}^{\pm}\non{\in} C_x^{1_{\pm}}
$
\\
\textbf{Location of the marginally stable orbits $r_{mso}$}
\\
\hline
\emph{The configurations} $\pp_1^{\pm}$
\\
$r_{mso}^{\pm}\in C_1^{\pm}$ $r_{mso}^{\pm}\in ! C_x^{1_{\pm}}$; \hspace{.5cm}$\forall \ell_1^{\pm}\in \mathbf{L1}^{\pm}$
\\
{\emph{The corotating  configurations} $\pp_2^-$
}\\
$a\in \breve{\mathbf{A}}_>$:  $ r_{mso}^-{\in}C_2^-\quad r_{mso}^-\in ! O_x^{2_-}$
\\
$a\in \breve{\mathbf{A}}_<$:  $\mathbf{I}\;r_{mso}^-\non{\in}C_2^-\quad r_{mso}^-\in ! O_x^{2_-}\quad \ell^-_2\in]\breve{\ell}_-, \ell_{\gamma}^-[\quad   r_{in}^{2_-}>r_{mso}^-$;\\
\hspace{1.3cm}   $
\mathbf{II}\; r_{mso}^-{\in}C_2^-\quad r_{mso}^-\in ! O_x^{2_-} \quad \ell_2^-\in]\ell_{mbo}^-, \breve{\ell}_-[$
\\
{{\emph{The counterrotating  configurations} $\pp_2^{+}$}}
\\
$ r_{mso}^+ \in! O_x^{2_+};\quad r_{mso}^+ \in C_2^{+}\quad \mbox{for}\quad
-\ell_2^+\in]-\ell^+_{mbo},-\breve{\ell}_+[$;\\
\hspace{2.2cm}  $r_{mso}^+ \non{\in} C_2^{+}\quad \mbox{for}\quad
-\ell_2^+\in]-\breve{\ell}_+,-\ell_{\gamma}^+[$
\\
\emph{The configurations} $\pp_3^{\pm}$
\\
$ r_{mso}^+\non{\in}C_3^+$
\\
$a\in \breve{\mathbf{A}}_<\quad r_{mso}^-\non{\in}C_3^-$;\\
$a\in \breve{\mathbf{A}}_>\quad r_{mso}^-{\in}C_3^-\quad \mbox{for}\quad
\ell_3^-\in]\ell_{\gamma}^-,\breve{\ell}_-[$, \;
$r_{mso}^-\non{\in}C_3^-\quad \mbox{for}\quad
\ell_3^->\breve{\ell}_-$
\\
\hline
\textbf{Location of the photon circular orbits $r_{\gamma}^{\pm}$}
\\
$r_{\gamma}^{\pm}\non{\in} \pp_{\pm}$
\\
\hline
\end{tabular}
\centering
\caption{Location of the notable radii $r_{\mathcal{N}}^{\pm} \in \pp_{\pm}$. This analysis sets location of the inner edge of the disk. In fact we have $r^{3_{\pm}}_{in}>r_{mbo}^{\pm} $; $r^{2_{\pm}}_{in}>r_{mbo}^{\pm}$, $r_{J}^{\pm}\leq r_{mbo}^{\pm};$ $r^{1_{\pm}}_{in}>r_{mbo}^{\pm} $ $r^{\pm}_{x}\geq r^{\pm}_{mbo}$--see also \citet{coop}  for further details on the classes of attractors.
}\label{Table:commest}
\end{table*}
\begin{table}[h!]
\begin{tabular}{ll}
Counterrotating configurations: $r_{\mathcal{N}}^{-}\in \pp_{+}$
\\\hline
\textbf{Location of the marginally bounded orbit $r_{mbo}^{-}\in \pp_{+}$}
\\
$r_{mbo}^-\non{\in} C_3^+$; \\
$a\in {\mathbf{A}}_{\iota}^>$: $r_{mbo}^-\non{\in} C_2^+$\;$r_{mbo}^-\non{\in} O_x^{2_+}$  \\
 $a\in {\mathbf{A}}_{\iota}^<:$ $r_{mbo}^-\non{\in} C_2^+$, $r_{mbo}^-\in O_x^{2_+} (r_{mbo}^-\succ r_J^{2_+}) \quad\mbox{for}\quad -\ell_+(r_{mbo}^-)\in[-\ell_{mbo}^+,-\ell_+(r_{mbo}^-)]$
\\
\hspace{1.4cm} $r_{mbo}^-\non{\in} C_2^+$, $r_{mbo}^-\non{\in} O_x^{2_+}\quad \mbox{for}\quad -\ell_+(r_{mbo}^-)\in[-\ell_+(r_{mbo}^-),-\ell_{\gamma}^+]$
\\
$r_{mbo}^-\non{\in} C_1^+\quad r_{mbo}^-\non{\in} C_x^{1_+}$
\\
\textbf{Location of the marginally stable orbits  $r_{mso}^{-}\in \pp_{+}$}
\\
\hline
\emph{The configurations} $\pp_1^{+}$
\\
$\mathbf{A}_{\iota_a}^>:\quad r_{mso}^-\non\in C_1^+ \; r_{mso}^-\non\in C_x^{1_+}$
 \\
 $\mathbf{A}_{\iota_a}^<:\quad  r_{mso}^-\non{\in}  C_x^{1_+}\; r_{mso}^-\non{\in} C_1^{+},\quad\mbox{for}\quad
 -\ell_1^+\in]-\ell_{mso}^+,-\ell_1^+(r_{mso}^-)[ $
 \\\hspace{1.1cm}
$r_{mso}^-\in C_x^{1_+} \;  r_{mso}^-\non{\in} C_1^{+}\quad\mbox{for}\quad \ell_1^+=\ell_1^+(r_{mso}^-),\quad  r_{in}^{1_+}=r_{Max}^{1_+}=r_{mso}^-=r_x^+ $
 \\\hspace{1.2cm}$
 r_{mso}^-\in! C_x^{1_+}\; r_{mso}^-\in  C_1^{+} \quad\mbox{for}\quad-\ell_1^+\in[-\ell_1^+(r_{mso}^-),-\ell_{mbo}^+[$
\\
\emph{The configurations} $\pp_2^+$
\\
$ r_{mso}^-\non{\in}C_2^+\quad \mbox{for}\quad-\ell_2^{+}\in]-\breve{\ell}_2^+,-\ell_{\gamma}^+[,$
\\
$ r_{mso}^-\non{\in}C_2^{+}\;
r_{mso}^-\non{\in}O_x^{2_+}\quad \mbox{for}\quad\quad a>a_{\gamma_+}^-\in \mathbf{A}^>_{\iota_a}$, $
r_{mso}^-\non{\in}C_2^{+} \quad \mbox{for}\quad a\in \mathbf{A}^>_{\iota_a},$
\\
$r_{mso}^-{\in}!O_x^{2_+}\quad \mbox{for}\quad a\in ]a_{\iota_a}, a_{\gamma_+}^-[\quad \mbox{and}\quad
-\ell_2^+\in[-\ell_2^+(r_{mso}^-),-\ell_{\gamma}^+[$
\\
$r_{mso}^-\in !O_x^{2_+} \quad \mbox{for}\quad a\in \mathbf{A}^<_{\iota_a}$, $\quad
r_{mso}^-{\in}C_2^{+} \quad \mbox{for}\quad a\in \mathbf{A}^<_{\iota_a} \quad \mbox{and}\quad -\ell_2^+\in]-\ell_{mbo}^+,-\breve{\ell}^-_{2_+}].$
\\
\emph{The configurations }$\pp_3^+$\\
$r_{mso}^-\non{\in}C_3^+$\\
\hline
\textbf{Location of the photon circular orbit $r_{\gamma}^-$}
\\
$r_{\gamma}^-\non{\in} \pp_+$
\\
\\
\emph{Corotating configurations}: $r_{\mathcal{N}}^{+}\in \pp_{-}$
\\\hline
\textbf{Location of the marginally stable orbits  $r_{mso}^{+}\in \pp_{-}$}
\\
$ r_{mso}^+\in C_-\quad\mbox{for}\quad a\in\breve{\mathbf{A}}^<_*\; \mathbf{L1}^-\cup[\ell_{mbo}^-,\breve{\ell}_*[\subset \textbf{L2}^-$,
\\\hspace{1.7cm} for $a\in\breve{\mathbf{A}}^>_*,\quad \mathbf{L1}^-\cup \mathbf{L2}^-\cup ]\ell_{\gamma}^-,\breve{\ell}_*[\subset \textbf{L3}^-$
\\
\textbf{Location of the marginally bounded orbits  $r_{mbo}^{+}\in \pp_{-}$}
\\
 $a\in\mathbf{A}^<_{\iota_a}$: $r_{mbo}^+\non{\in}C_3^-;\quad r_{mbo}^+\non{\in}C_2^-\quad \mbox{for}\quad \ell^-_2\in]\breve{\ell}_-, \ell_{\gamma}^-[\quad r_{in}^{2_-}>r_{mso}^->r_{mbo}^+.$
 \\
$a>a_{\gamma_-}^{\beta}:$ $r_{mbo}^+\in \pp^-_1;\; r_{mbo}^+\in \pp^-_2 $; $r_{mbo}^+\non{\in} C^-_3\quad\mbox{for}$ $\ell>\ell_{\beta}^{-}$; \; $r_{mbo}^+{\in} C^-_3$ for $\ell<\ell_{\beta}^{-}$
 \\
 $ a<a_{\gamma_-}^{\beta}:$ $r_{mbo}^+\in \pp^-_1;\; r_{mbo}^+\in \pp^-_2 $ for  $\ell_-<\ell_{\beta}^-$;
 $\; r_{mbo}^+\non{\in} \pp^-_2\quad\mbox{for}\quad
\;\ell_->\ell_{\beta}^-$;  $r_{mbo}^+\non{\in} C^-_3$
 \\
 \hline
 \textbf{Location of the photon circular orbit $r_{\gamma}^+$}
 \\
$a\in[0,a_{\gamma_-}^{\Gamma}[:$
   $r_{\gamma}^+\in \pp^-_1 $  $r_{\gamma}^+\non{\in}C_3^-$;
  $ r_{\gamma}^+\in \pp^-_2 $ for $ \ell_-<\ell_{\Gamma}^-$; $r_{\gamma}^+\non{\in }\pp^-_2$ for $ \ell_->\ell_{\Gamma}^-\in \mathbf{L2}^-$
   \\
   $a\in]a_{\gamma_-}^{\Gamma},M]^*:$ $
r_{\gamma}^+\in \pp^-_1\; r_{\gamma}^+\in \pp^-_2$;
$ r_{\gamma}^+\in \pp_3^-$ for $\ell_-<\ell_{\Gamma}^-\in \mathbf{L3}^-$; $r_{\gamma}^+\non{\in }\pp_3^-$ for $ \ell_->\ell_{\Gamma}^-\in \mathbf{L3}^-$.
\\
\hline
\end{tabular}
\centering
\caption{Location of the notable radii $r_{\mathcal{N}}^{\pm} \in \pp_{\mp}$. This analysis sets location of the  edge of the disk.
}\label{Table:commestmp}
\end{table}
\textbf{Table\il(\ref{Table:commetime})} fixes the five seed states   with at least one cusped topology and the possible  correlation, establishing also the  critical  and centers   sequentiality, i.e.,  the relative location of  rings of a couple in a particular geometry.
We introduce the spin classes:
\bea
&&\mathbf{A}_{\iota}^<\equiv[0,a_{\iota}[\quad \mbox{and}\quad\mathbf{A}_{\iota}^>\equiv[a_{\iota}, M] \quad \mbox{where}\quad a_{\iota}\equiv0.3137M:r_{mbo}^-=r_{\gamma}^+,
\\
&&
\mathbf{A}_{\iota_a}^<\equiv[0,a_{\iota_a}[\quad \mbox{and}\quad \mathbf{A}_{\iota_a}^>\equiv[a_{\iota_a}, M]\quad \mbox{where}\quad a_{\iota_a}\equiv0.372583M:r_{mso}^-= r_{mbo}^+.
\eea
Some states    are uniquely fixed, some states are not possible. The correlation is possible in all corotating couples, according to a proper choice of the ``density'' $K$-parameter.   The ring separation induced on the  $\ell$counterrotating couples,
due to the attractor rotation,  acts  in general  to forbid or  to disadvantage the correlation--see also \citet{ringed}. This situation is obviously  less clear as the spin of the attractor decreases. This would imply that the   $ \ell$counterrotating ring  dynamics  can be often  described as  collection of independent (separate) configurations.
A more detailed  look of Table\il(\ref{Table:commetime}) reveals that in the  $\ell$counterrotating case the angular momentum is not sufficient to uniquely fix the state and its correlation, but the situation  depends on the dimensionless spin of the attractor.
In conclusion, the {proto jet-accretion} correlation  is always possible, except in the $\ell$counterrotating  couples  in the geometries of the fastest attractors\footnote{For \emph{fast} (\emph{slow}) attractors we intend Kerr attractors with  high ({small}) values of the dimensionless spin with  respect to some  reference values of $a/M$, fixed considering the  geodesic structure  of the spacetime  in Fig.\il(\ref{Figs:PlotdisolMsM})).}, $\mathbf{A^>_{\iota_a}}$, with corotating fluid in accretion and counterrotating proto-jet  where the point of accretion is always inner with respect to $r_J$.
The case of accretion of counterrotating matter  and corotating proto-jet is favored, according to the constraints, in any Kerr geometry,  while  the case of corotating accretion disk, and counterrotating matter with launching point inner  or coincident with the accretion point  is favored in the geometries of  the slow
attractors,  $\mathbf{A_{\iota_a}^<}$.
A {$O_x$-$O_x$} correlation (i.e. a double-shell of open configurations in contact), is always possible except for the fast attractors, $\mathbf{A_{\iota}^>}$, where  it is prohibited for an outer counterrotating proto-jet;
it is possible  for any  combination of launching  points at low spin  geometries   $\mathbf{A_{\iota}^<}$.
The {accretion-accretion} correlation is possible  but  with an inner corotating accretion point.
The  {accretion-equilibrium}  correlation is impossible or
	 not favored for attractors with large spins $\mathbf{A_{\iota_a}^>}$  (being restrictive conditions on the ring parameters) for   the counterrotating disk, which must be the outer of the couple with respect to corotating accreting disk. In the case of corotating equilibrium disks a correlation is possible.
The  {proto jet-equilibrium}  correlation is not possible if  there is a corotating  jproto-et, and
 not favored, if the corotating   equilibrium disk is outer with respect to the proto-jet configuration.
\textbf{Table\il(\ref{Table:commest})}
  indicates the location of the orbiting  configurations with respect to the $\ell$corotating geodesic structure of the Kerr  geometry, where there is
\bea
&&\mathbf{\breve{A}}^<\equiv[0,\breve{a}[\quad \mbox{and}\quad\mathbf{\breve{A}}^>\equiv[\breve{a}, M] \quad \mbox{with}\quad \breve{a}\equiv 0.969174M:\,\breve{\ell}_-=r_{\gamma}^-
\\
&&
\mbox{and}\quad \breve{\ell}_{\pm}(a/M):\; V_{eff}(\breve{\ell}_{\pm},r_{mso}^{\pm})=1.
\eea
%
 We analyzed the inclusion relations  $r_{\mathcal{N}}^{\pm} \in \pp_{\pm}$ setting, particularly, the location of the ring inner edge.
The {marginally bound orbit} is never included in any equilibrium  or accretion topology. This implies that the disk inner edge    must be always external to this, whereas the launching point of proto-jet must be internal.
The { marginally stable orbit} must be always included in any unstable ring and it can   also be included  in the  equilibrium  $C^{\pm}_1$  configuration  with a  density lower then the critical one ($K<K_{Max}$).
The crossing  of the marginally stable orbit does not lead, by itself, to the P-W instability.
 The situation at higher spin is  much more complex and depends generally on two classes of attractors and the direction of rotation of the fluid with respect to these. For corotating disks in $\mathbf{L2}$, whose unstable mode is a proto-jet, orbiting fast attractors,  $\mathbf{\breve{A}_>}$, the marginally stable orbit can be included in their equilibrium configurations ($r_{in}<r_{mso}^-$ depending on the  $K$-parameter). At lower spins, $\mathbf{\breve{A}_<}$, this cannot occur (as $r_{in}>r_{mso}^-$) but for  lower  specific angular momentum.
 For the counterrotating configurations the occurrence of $r_{mso}^+\in C_2^+$ does not depend directly  on the  attractors but  again for high magnitudes of specific  angular momentum, the marginally stable orbit is not included in the disk, the limiting specific angular momenta being however function of the dimensionless $a/M$\citep{coop}.
At higher specific angular momentum magnitudes, $\mathbf{L3}$,  for which there is no unstable mode, there has to be   $r_{in}^+>r_{mso}^+$ for any attractor. Whereas $r_{in}^->r_{mso}^-$  for $\mathbf{\breve{A}_<}$, while in $\mathbf{\breve{A}_>}$ the marginally stable orbit may be included in $C_3^-$ for low specific angular momentum.
\textbf{Table\il(\ref{Table:commestmp})} indicates the location of  each ring with  respect to the $\ell$counterrotating geodesic structure of the Kerr  geometry, examining the inclusion   $r_{\mathcal{N}}^{\mp} \in \pp_{\pm}$, where  there is
\bea
&&\mathbf{\breve{A}^<_*}\equiv[0,\breve{a}_*[\; \mbox{and}\;\mathbf{\breve{A}^>_*}\equiv[\breve{a}_*, M] \;\;\mbox{where}\;\;\breve{a}_*\equiv 0.401642 M:\breve{\ell}_*=\ell_{\gamma}^-,\,
\mbox{and}\;\; \breve{\ell}_*: V_{eff}(\breve{\ell}_*,r_{mso}^+)=1,
\\
&&
a_{\gamma_-}^{\beta}\equiv0.628201 M:\ell_{\beta}^-=\ell_{\gamma}^-\quad \mbox{where}\quad {\ell}_{\beta}^-:\; V_{eff}(\ell_{\beta}^-,r_{mbo}^+)<1,
\\
&&a_{\gamma_-}^{\Gamma}\equiv0.777271M: \ell_{\Gamma}^-=\ell_{\gamma}^-\quad \mbox{where}\quad  \ell_{\Gamma}^{-}:\; V_{eff}(\ell_{\Gamma}^{-},r_{\gamma}^+)=1,
\\
&&a_{\gamma_+}^-\equiv 0.638285 M:r_{\gamma}^+=r_{mso}^-,
\\
&&\breve{\ell}^-_{2_+}:\;V_{eff}(\ell_2^+, r_{mso}^-)=1\quad\mbox{and}\quad \breve{\ell}^-_{2}=\breve{\ell}_-\in \mathbf{L2}.
\eea
This analysis provides also  more restrictive constraints  for  the location of the inner edge of a ring, providing  also information
on the  ring outer margins.

The study of the inclusion of the ring margins with respect to  the $\ell$counterrotating geodesic structure of the  Kerr geometry  is essential to establish a possible correlation between the
  $\ell$counterrotating fluids and related  evolution. Several results of this analysis were in fact used in  Table\il(\ref{Table:commetime})--see also \citet{coop}.
We  focus  first on the counterrotating fluids. According to  Table\il(\ref{Table:commestmp}),   the  proto-jet launching point  can be internal to $r_{mbo}^-$  for sufficiently low  spin, $\mathbf{A_{\iota}^<}$,  and low specific  angular  momenta in magnitude only. 
Then the launching point must be external to $r_{mso}^-$ in the geometries  $a>a_{\gamma_+}^-$.
 It must be internal for slower attractors, and  in  $\mathbf{A_{\iota_a}^>}$    for sufficiently high angular momentum  magnitudes.
It  must be internal for low spins,  $\mathbf{A_{\iota_a}^<}$,  and for all values of specific angular momentum.

We focus now on  the inner edge of the counterrotating disk in equilibrium or in accretion with respect to $r_{mso}$. At this point the analysis is more  complicated  because the double geodesic structure includes crossing  points between the elements of $r_{\mathcal{N}}^{\pm}$, therefore the situation  strongly depends on the geometric properties of the Kerr spacetimes.
We show that the   ratio $\ell/a$ must  have specific characteristics for an instability to emerge.
Concerning  the accretion points,
the most significant aspect perhaps is that the point of accretion must be internal with respect to  $r_{mso}^+$-- Table\il(\ref{Table:commest}), but  ring must not include $r_{mso}^-$ in the $\mathbf{A_{\iota_a}^>}$ geometries-- Table\il(\ref{Table:commestmp}).  Viceversa, we have to take into account if the  fluid is orbiting $\mathbf{A_{\iota_a}^<}$ attractors with sufficiently high  specific momentum magnitude.
On the other hand, focusing on the   $C^+_1$ disk,  this can contain  the orbit $r_{mso}^+$ without being unstable,
but it cannot contain $r_{mbo}^-$ for any \textbf{BH} spin; it cannot contain $r_{mso}^-$ for  $\mathbf{A_{\iota_a}^>}$ class,  it can be included viceversa for  sufficiently slower spin attractors, $\mathbf{A_{\iota_a}^<}$ and large magnitude of the specific angular momentum.
 The inner edge  of the  equilibrium  $C_2^+$ configurations   can be internal to $r_{mso}^-$
only for  slow attractors and specific angular  momentum with  a sufficiently low magnitude.

The more articulated situation  concerns  the corotating disks and their location with respect to the counterrotating geodesic structure.
For this analysis  it was necessary to thoroughly analyze  the situation of the outer edge of the closed topology, as there are in fact $r_a^-<r_a^+$ for  $r_a\in r_{\mathcal{N}}$. Here we deal  the case $r_a^+\in[r_{in}^-,r_{out}^-]$.
We start with the inclusion  of the photon orbit $r_{\gamma}^+$ which is  the innermost radius   in the structure determined by $r_{\mathcal{N}}^{+}$.
There are two classes of attractors. For  $a<a_{\gamma_-}^{\Gamma}$, radius $r_{\gamma}^+$ can always be  in  $\pp^-_1$, but  never in $C^-_3$,
while for  $r_{\gamma}^+\in\pp^-_2$ a sufficiently low angular momentum is required.
  For faster attractors,  there is in general always    $r_{\gamma}^+\in\pp^-_1$ and  $\pp^-_2$,
  but  in $ C^-_3$ a sufficiently low specific  angular momentum is required.
   The orbit $r_{mbo}^+\in C^-_1$, but is never in $C^-_3$ for $a\in\mathbf{A^<_{\iota_a}}$,
   in this class it is in $C^-_2$ only for sufficiently  low specific angular  momentum. Al larger $a/M$ it is always $r_{mbo}^+\in C^-_2$, while there is  $r_{mbo}^+\in C^-_3$ only for  the lower $\ell_3^-$.
   The situation is in general more articulated for  $ C^-_2$ and  $C^-_3$ and it depends on   $a_{\gamma_-}^{\beta}$: for large $a/M $ there is $r_{mbo}^+\in \pp_2^-$ only for large specific momenta,
for large spin there can be always   $r_{mbo}^+\in \pp_2^-$  while it is in $C_3^-$  only for low enough momenta.
The marginally stable orbit $r_{mso}^+>r_{mso}^-$  can always be in $\pp_1^-$ for the geometries $\mathbf{\breve{A}_*^<}$,  and in $\pp_2^-$ only for  low specific angular momentum.  For large  spin attractors  instead it can be included in $\pp_1^-$  and  $\pp_2^-$, and  in $\pp_3^-$  only for lower spins.
%
Further discussion on methods and applications of the instabilities,   details of each individual state and the precise definition of the limits of angular momentum will be presented in \citet{coop} where it is  also shown  how
composition of more rings in  macro configuration  is in many cases strongly constrained and evolution limited.
\section{Conclusions and Future Perspectives}\label{Sec:Open-Concl}
We  addressed   instabilities emerging in the ringed  accretion disks introduced in \citet{ringed}, considering    in  particular conditions for  the emergence of  unstable phases for each  ring   according to the  P-W mechanism and the condition for the destabilization  of macro-configuration  after a rings collision.
 First, we proved that the existence of a couple of rings is strongly constrained with respect to  the  relative location of the configurations in the couple, the ring instability, the relative rotation, and the rotation with respect to the attractor and, in the case of $\ell$counterrotating rings, the dimensionless spin on the central Kerr black hole.
Considering  ringed disks  with at last one critical point, we established  the   orbital and critical  rings sequentiality, the
location of instability points, and the  inner and outer margins of each ring, as well as the    emergence of possible  contact points between two sub-configurations (correlation), which can  lead to collision in a  $\mathbf{C}_{\odot}^n$ unstable  macro-configuration--Table\il(\ref{Table:commetime}).
Some   results of this investigation   naturally apply    to the case of only one accretion disk, which  can be considered as the simplest   ringed  disk  of the  order  $n=1$. The analysis, in the  framework of the ringed disks, in particular   th clarifies   significant and widely debated issue of the location of the inner  (and outer) edge of  an accretion disk, where in general     emergence of P-W instability occurs. We  constrained the location of the inner and outer edges of each   toroidal ring   according to its  specific angular momentum, and   the geodesics structure of the Kerr spacetime, and the $K$-parameter involved in the
ring density definition--Tables\il(\ref{Table:commest},\ref{Table:commestmp}).
This study was in fact relevant  in the  determination of rings relative location  and the recognition of correlation.
Properties of the ringed disk, reflected  in Tables\il(\ref{Table:commetime},\ref{Table:commest},\ref{Table:commestmp}), could be used to identify  the background geometry in one of the  particular class of attractors  considered here.
 Highlighting  the possibility that there may be structures  formed by more then one accretion torus  orbiting around a Kerr attractor we believe that present work could  have significance in the high-energy astrophysics, playing  possibly a significant role in some of  already observed phenomena, namely in terms of interaction and  dynamics of several rings  of the macro-configuration.
The  unstable modes and the ring collisions  could  lead to several  phenomena  eventually involving  the attractor itself, like  for example the  runaway instability
 \citep{Abra83,Abramowicz:1997sg,Rez-Zan-Fon:2003:ASTRA:,Font:2002bi,Hamersky:2013cza,Lot2013}.
From methodological point of view, by considering a  purely hydrodynamic model with  a  constant specific  angular momentum,  this analysis captures some  significant aspects of the basic geometric properties
of the     extended matter configurations, considered in some way as the relativistic generalization  of  geodesic structure to the extended objects with pressure gradients, relevant for this structure.
Thus, the considerations traced here  could be applicable  also for more general   models where
the specific angular  momentum is not constant along the disk  \citet{Lei:2008ui}.
A further generalization of this work  should  consider the role of the magnetic fields in the macro-structures.
\acknowledgments
The authors acknowledge the institutional support of the   Faculty of Philosophy and Science of the Silesian University of Opava.
Z.S.  acknowledges the Albert Einstein Center for Gravitation and Astrophysics supported by the Czech Science Foundation grant No. 14-37086G.   D. P. acknowledges also the Accademia Nazionale dei Lincei for support during 2015 (within the Royal Society fellowship program),  the Junior GACR grant of the Czech Science Foundation No:16-03564Y, and useful discussions  with  Prof. J. Miller, Prof. M. A. Abramowicz, Prof. V. Karas  at RagTime17 workshop.

\end{document}